\newcommand{\CLASSINPUTtoptextmargin}{0.75in}
\newcommand{\CLASSINPUTbottomtextmargin}{1.1in}
\documentclass[conference]{IEEEtran}
\IEEEoverridecommandlockouts
\usepackage{cite}
\usepackage{amsmath,amssymb,amsfonts}
\usepackage{graphicx}
\usepackage{textcomp}
\usepackage{booktabs, multirow, tabularx}
\usepackage{amsmath,amsfonts}
\usepackage{algorithm}
\usepackage{algpseudocode}
\usepackage{amsmath,amsfonts,amssymb}
\usepackage{tabularx}
\usepackage{array}
\usepackage{amsthm}
\usepackage{array}
\usepackage{textcomp}
\usepackage{stfloats}
\usepackage{url}
\usepackage[table]{xcolor}

\def\BibTeX{{\rm B\kern-.05em{\sc i\kern-.025em b}\kern-.08em
    T\kern-.1667em\lower.7ex\hbox{E}\kern-.125emX}}

\makeatletter

\let\old@ps@IEEEtitlepagestyle\ps@IEEEtitlepagestyle

\def\confheader#1{%
    \def\ps@IEEEtitlepagestyle{%
        \old@ps@IEEEtitlepagestyle%
        \def\@oddhead{\strut\hfill#1\hfill\strut}%
        \def\@evenhead{\strut\hfill#1\hfill\strut}%
    }%
    \ps@headings%
}
    
\makeatother
    \confheader{%
    \parbox{20cm}{Accepted by 2024 IEEE 10th World Forum on Internet of Things, (IEEE WF-IoT), \textcopyright 2024 IEEE}
}

\begin{document}

\title{Q-CSM: \textbf{Q}-Learning-based \textbf{C}ognitive \textbf{S}ervice \textbf{M}anagement in Heterogeneous IoT Networks\\
}

\author{\IEEEauthorblockN{Kubra Duran\IEEEauthorrefmark{1},
Mehmet Ozdem\IEEEauthorrefmark{3}, Kerem Gursu\IEEEauthorrefmark{4}, Berk Canberk\IEEEauthorrefmark{1}\IEEEauthorrefmark{2}}
 \
\IEEEauthorblockA{\IEEEauthorrefmark{1}School of Computing, Engineering and The Built Environment, Edinburgh Napier University, UK}
\IEEEauthorblockA{\IEEEauthorrefmark{2}Department of Artificial Intelligence and Data Engineering, Istanbul Technical University, Türkiye}
\IEEEauthorblockA{\IEEEauthorrefmark{3}Turk Telekom, Istanbul, Türkiye}
\IEEEauthorblockA{\IEEEauthorrefmark{4}BTS Group, Istanbul, Türkiye}

		Emails: \{kubra.duran, b.canberk\}@napier.ac.uk, mehmet.ozdem@turktelekom.com.tr, kerem.gursu@btsgrp.com}

\maketitle

\begin{abstract}
The dramatic increase in the number of smart services and their diversity poses a significant challenge in Internet of Things (IoT) networks: heterogeneity. This causes significant quality of service (QoS) degradation in IoT networks. In addition, the constraints of IoT devices in terms of computational capability and energy resources add extra complexity to this. However, the current studies remain insufficient to solve this problem due to the lack of cognitive action recommendations. Therefore, we propose a Q-learning-based Cognitive Service Management framework called Q-CSM. In this framework, we first design an IoT Agent Manager to handle the heterogeneity in data formats. After that, we design a Q-learning-based recommendation engine to optimize the devices' lifetime according to the predicted QoS behaviour of the changing IoT network scenarios. We apply the proposed cognitive management to a smart city scenario consisting of three specific services: wind turbines, solar panels, and transportation systems. We note that our proposed cognitive method achieves 38.7\% faster response time to the dynamical IoT changes in topology. Furthermore, the proposed framework achieves 19.8\% longer lifetime on average for constrained IoT devices thanks to its Q-learning-based cognitive decision capability. In addition, we explore the most successive learning rate value in the Q-learning run through the exploration and exploitation phases. 

\end{abstract}


\section{Introduction}
In recent years, the integration of Internet of Things (IoT) technologies into urban infrastructures has revolutionized the concept of smart cities. IoT sensors, characterized by their small compute footprint and described as constrained nodes, play a pivotal role in this transformation \cite{survey2}. These sensors are constrained by limited computational power, storage, and energy resources, which pose significant challenges in terms of the complexity of tasks that the devices can handle\cite{tgcn}. This emphasizes the necessity for intelligent and resource-efficient management solutions in constrained networks where devices must operate persistently and autonomously by serving the desired Quality of Service (QoS) levels\cite{intelligent}. 

According to RFC7228 \cite{rfc7228}, IoT devices are categorized into three distinct types of constrained sensors. When these types are present in a single network, it becomes a heterogeneous IoT sensor network in the environment. At this point, this heterogeneity introduces significant complexities in the deployment and management of real-world smart cities due to its specific requirements \cite{dtaas}. In this study, we deal with the heterogeneity problem from two perspectives:

\begin{itemize}
    \item \textit{Heterogeneity in terms of IoT devices:} Divergent types of IoT devices draw a picture of different hardware and software at the backend. These differences require several types of communication protocol implementation depending on their connection types \cite{bozkaya}. In addition to this, different configuration methods should be leveraged at this scenario.
    \item \textit{Heterogeneity in terms of QoS:}  The functionality of the IoT sensors varies from simple sensing to high-processing devices. While some IoT nodes perform monitoring tasks, others might require high-definition information. Such a vast variation in the requirements of IoT nodes translates to significantly different QoS expectations for IoT networks.
\end{itemize}

\section{State of the Art}
Dynamic conditions in heterogeneous IoT networks pose a significant challenge to communication quality. For this reason, the current literature covers this problem from a resource-aware service \cite{ccnc} perspective. \cite{contextaware} proposes context-aware connectivity and processing optimization in IoT networks. In this study, the joint optimization of energy consumption and response time is considered by the Reinforcement Learning (RL) algorithm. Furthermore, \cite{protocoladaptive} and \cite{camad} introduce a protocol-adaptive Software Defined Networks (SDN)-based solution for the dynamic network conditions in smart city applications. Similarly, \cite{access2} and \cite{icc24} introduce an RL-based solution for the resource allocation and communication delay problems in IoT networks. Also, a fog-layered service management scheme is designed for IoT-based smart cities in \cite{trconsumer}. As IoT applications have several resource constraints and are sensitive to latency and dynamical changes, the QoS level of these services is affected by these \cite{wiley}. For instance, \cite{schedular} performs context-aware fog computing for an effective load-balancing scheme. With this, the study leverages context-sharing, context-migration, and live service migration strategies based on the prediction algorithms. Furthermore, advanced AI models are integrated into smart city applications to ease QoS management. For instance, \cite{access} introduces the application of generative adversarial networks (GANs) to smart city applications to generate synthetic data. Another study, \cite{IoT_journal}, focuses on service quality from an infrastructural perspective to meet the desired quality levels. Besides, the autonomic applications term for IoT is presented in \cite{survey} by surveying the quality metrics for IoT networks. This study focuses on real-time diagnostics of heterogeneous IoT applications and making them quality-aware. 

\begin{table*}[!t] 
\centering
\caption{Proposed Q-CSM Framework and Current State of the Art Studies}
\label{tab:lit}
\begin{tabular}{l|ccccc}
\toprule
\hline
\multicolumn{1}{c|}{\textbf{Literature}} & \textbf{\begin{tabular}[c]{@{}c@{}}Heterogeneity\end{tabular}} & \textbf{QoS-aware} & \textbf{\begin{tabular}[c]{@{}c@{}}Intelligence\end{tabular}} & \textbf{\begin{tabular}[c]{@{}c@{}}Prediction method\end{tabular}} & \textbf{\begin{tabular}[c]{@{}c@{}}Cognitive actions\end{tabular}} \\ \hline
\cite{contextaware}, \cite{protocoladaptive}, \cite{camad}, \cite{access2}, \cite{icc24}               & \checkmark                        & \checkmark             & \checkmark                          & RL                            & -                            \\ 
\cite{trconsumer}, \cite{schedular} & \checkmark                        & -                 & \checkmark                        & -        & -                        \\ 
\cite{access}, \cite{IoT_journal}                               & \checkmark            & -                     & \checkmark                                                                        & GAN                                                                  &-                                             \\
\cite{survey}                                 & \checkmark            & \checkmark              & -                                                                        & -                                                                    & -                                            \\
Our work                                 & \checkmark             & \checkmark                  & \checkmark                                                                         & Q-Learning                                                                   & \checkmark                                        \\ \hline
\bottomrule
\end{tabular}
\end{table*}

As summarized above, several efforts have been made to address the communication latency and QoS challenges stemming from the heterogeneity of IoT networks. Although some of them apply AI methods to serve intelligence in management, they only utilize the algorithms to predict the resource usage rate and possible faults on the network. None of them covers exploring the IoT topology to learn its behaviour as a whole and create autonomous actions, which stand as a cognitive management framework. Therefore, our research is situated around the research question, 
\textit{“How can we maintain multiple classes of IoT devices within a single smart city network and meet the particular QoS requirements by optimizing the service response time and increasing the IoT devices' lifetime?”} To hit this, we propose a Q-learning-based Cognitive Service Management framework, called as Q-CSM. In Q-CSM, we model three distinct layers to manage the heterogeneity problem efficiently. We first consider the constrained IoT device classes and form and adaptation layer to manage the constrained classes within the single IoT topology. Afterwards, we perform a Q-learning algorithm to produce optimized actions regarding the lifetime of IoT devices and desired QoS levels. Our main goal is to dynamically configure IoT device communication to meet QoS requirements in changing IoT network scenarios. The main contributions of this study are summarized below:

\begin{itemize}
    \item We design an IoT Agent Manager within the Adaptation Layer to transform all data types to a single data format and thus serve as a multi-tenant gateway in heterogeneous IoT networks.
    \item We design a Q-learning-based recommendation engine to optimize the devices' lifetime by serving device management according to the predicted QoS behaviour of the changing IoT network scenarios. 
    \item We apply the proposed cognitive management on a smart city scenario consisting of three specific services: wind turbines, solar panels, and transportation system.
\end{itemize}

The remainder of the article is organized as follows: Section
III explains the proposed cognitive service management framework. Section IV is devoted to the performance evaluation of the proposed model. Finally,
Section V finalizes the paper.

\begin{figure*}[b]
\centerline{\includegraphics[width=.63\textwidth]{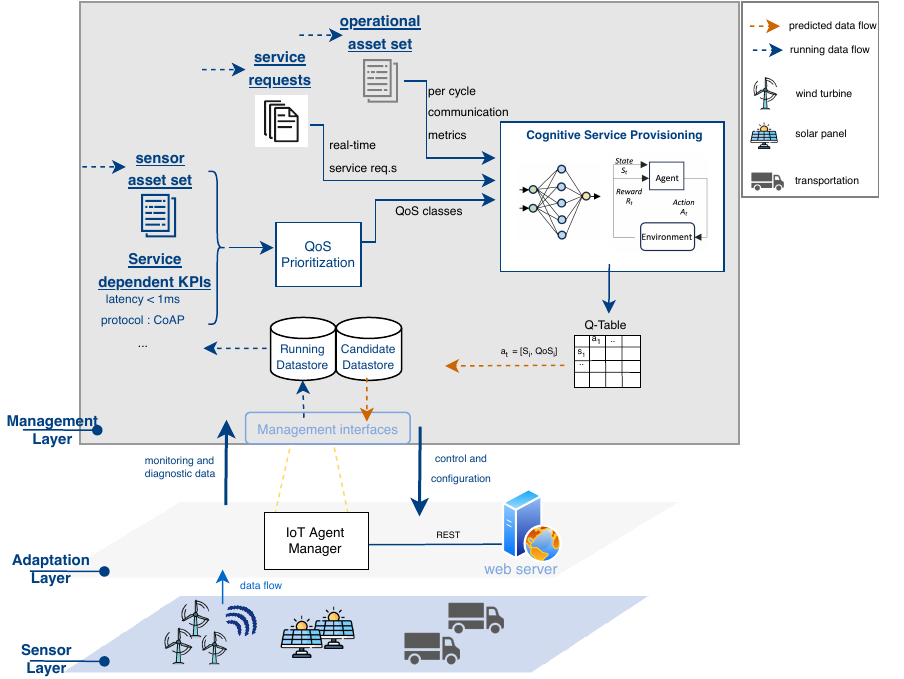}}
\caption{Proposed cognitive service management framework.}
\label{bigpicture}
\end{figure*}

\section{Q-CSM: Cognitive Service Management Framework}
As given in Fig. \ref{bigpicture}, our proposed cognitive management framework consists of three layers; \textit{Sensor Layer}, \textit{Adaptation Layer} and \textit{Management Layer}. The details of these layers are explained in below sections.

\subsection{Sensor Layer}
This layer consists of deployed IoT sensors for smart city scenarios. To cover the heterogeneity problem form IoT devices perspective, we assume that the IoT sensors in this layer embody all types of constrained device classes with the battery-powered feature. The specifications for these constrained device classes are explained below:
\begin{itemize}
    \item \textit{Class 0:} This class is strictly constrained in terms of memory and processing capabilities. For example, the maximum data size to be supported 
    \item \textit{Class 1:} This class is quite constrained regarding the code space and processing capabilities. 
    \item \textit{Class 2:} This class is less constrained compared to Class 0 and Class 1. Also, this class is capable of supporting protocol stacks utilized in servers.
\end{itemize}



\subsection{Adaptation Layer}
\begin{figure}[h]
\centerline{\includegraphics[width=.36\textwidth]{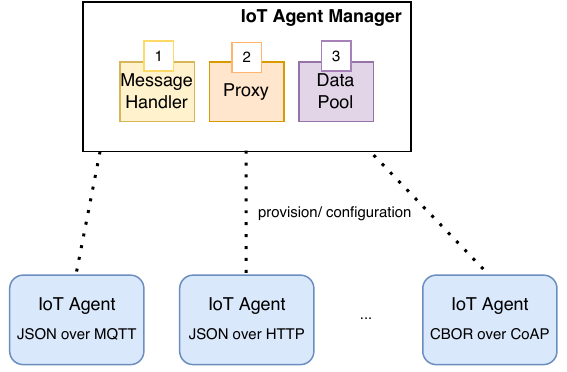}}
\caption{Design of IoT Device Manager.}
\label{adapt}
\end{figure}

This layer collects sensor data and serves as a proxy between the sensor layer and the management layer. IoT application layer protocols run on constrained nodes with a small compute footprint \cite{aot}. As Message Queuing Telemetry Transport (MQTT), Constrained Application Protocol (CoAP), and Hypertext Transfer Protocol (HTTP) are mainly utilized within the application layer of IoT protocol stack, we assume three of them are implemented within the IoT agents. As given in Fig. \ref{adapt}, IoT Agent Manager comprises three sub-modules, which are explained below:

\begin{itemize}
    \item \textit{Message Handler:} Messages coming from IoT agents first arrive here. As the IoT Device Manager implements MQTT, HTTP and CoAP, both Transport Control Protocol (TCP and User Datagram Protocol (UDP) are utilized in the transport layer regardless of the wireless protocol in the lower layers. After that, collected data in the form of JSON and CBOR is sent to the next module.
    \item \textit{Proxy:} It stands between the message handler and data pool and behaves as a tunnel. The main role of this module is to translate the data formats to JavaScript Object Notation (JSON) to be processed within the management layer. Namely, this module transforms all the CBOR data types into JSON format. The translation process consists of encoding and decoding of multiplexers.
    \item \textit{Data Pool:} The data flowing through the proxy come to this module. Data pool stores the collected application data from all IoT agents in the form of JSON. This is because, the IoT network data is collected via the management interfaces of the management layer in this format.  
\end{itemize}

\subsection{Management Layer}
The main role of this layer is to perform an intelligent management of different services within a single IoT network. For this, first the data is collected from the adaptation layer via its management interfaces. Afterwards, it is sent to the \textit{running datastore}, where all the records at the time interval $[t-x, t]$ are stored. Here, $t$ is in seconds, and $x$ is the user-changeable parameter depending on the application scenario and the capacity of the utilized database. Conversely, the \textit{candidate datastore} holds the recommended actions that are outputted by the Q-learning algorithm. 

\begin{table}[h]
\caption{Considered Smart City Scenarios and Respective KPIS}
\centering
\begin{tabular}{l c c c}
\hline
No. & Scenario & KPI Specs & Protocol\\
\hline
$1$ & Wind Turbine & delay $\leq$ 300ms, loss rate $\leq$ 10\% & CoAP\\
$2$ & Solar Panel & delay $\leq$ 300ms, loss rate $\leq$ 10\% & HTTP\\  
$3$ & Transportation & delay $\leq$ 100ms, loss rate $\leq$ 5\% & MQTT\\
\hline
\end{tabular}
\label{tab:table2}
\end{table}

One of the main steps of this layer is to perform QoS prioritization regarding the smart city services. As a single, smart city network comprises several different services \cite{t6conf}, thus they compromise to different QoS requirements depending on their service target. Therefore, service dependent KPIs \cite{wifi} along with the sensor asset set are considered to form the QoS levels of the smart city services. In Table I, we give three specific smart city scenarios that we consider and define their communication requirements with the utilized protocol at the application layer.
We process the QoS class formation by considering the total number of active IoT devices in the running environment. Therefore, we calculate each of the QoS densities as, 

\begin{equation}
\alpha(Q_i)=\frac{\sum_{m} O_i}{V_{Q_i}}\label{eq1}
\end{equation}

where \[1 \leq i \leq 2\ \text{,}\  1 \leq m \leq n\]

In this formula, $\alpha(Q_i)$ is the $i^{th}$ QoS class density, $O_i$ is the total number of active IoT devices, and $V_{Q_i}$ implies the number of active IoT devices waiting in the related queue to take service from the IoT Manager. Also, $i$ is the number of QoS classes, and $m$ is the total streamed number of IoT devices. Also, we adapt the Q-Learning to create actions. We map the Q-learning components into our smart city IoT topology by considering that there is one master node for each of the smart city service, and this master is an agent. Also, we consider each QoS class change to be a state within the IoT scenario. The agent performs action by exploring the optimum QoS class for the smart service and obtains a reward for this. Therefore, our cognitive engine is represented as six tuples; \textit{\{$E, Ag, S, A, p, R$\}}. We explain the major elements of this system below.

\begin{itemize}
    \item \textit{Environment, $E$:} The smart city network consisting of $n$ number of IoT sensors forms an environment for Q-learning. The smart city network consists of three services: wind turbines, solar panels and transportation.
    \item \textit{Agent, $Ag$:} Each service in the smart city network has a master node to behave as an agent. The agent is capable of exploring the simulated environment for learning phase. 
    \item \textit{State, $S$}: The state space represents the total number of QoS classes depending on the defined service specific KPIs and utilized application layer protocol.
    \item \textit{Action, $A$:} Action set states which smart city service should be served with which QoS class depending on the real-time service requests, historical records, and the operational asset set.
    \item \textit{Probability, $p$:} Represents the probability of transition to the new state. In our approach, it stands for the probability of transitioning from one QoS class to another.
    \item \textit{Reward, $R$:} 
    The agents accept a reward for each QoS class change in which the requested KPIs are optimized considering the whole smart city network.
    \item \textit{Q Function:} The algorithm updates the Q values by using the equation which we adapt from Bellman Equation: \(Q_{t+1}(S, A)= Q(.) + \alpha (R + \gamma \max_{A'} Q(S',A')-Q(.))\), where $Q(.)$ function stands for $Q_{t}(S, A)$. In this formula, the $A'$ is the action that could be taken at state $S'$. Also, $\gamma$ is a discount factor showing the significance of next states with learning rate, $\alpha$.  

 \item \textit{Policy, $\pi$:}   We utilize the $\epsilon$-greedy  action selection mechanism by randomly choosing the exploration and exploitation states. Therefore, the agent takes a random action at a given time with the probability of $\epsilon$ or (1 $-$$\epsilon$).
\end{itemize}

As seen in Alg.1, Q-CSM takes the number of smart city services, number of active IoT devices, KPI specifications, sensor asset set, operational asset set, and service request information as inputs and produces QoS classes, QoS class density values, and optimized actions to increase the average lifetime of IoT devices. In line-1, the density of the QoS classes is initialized before starting the first run of the algorithm. When the number of active IoT devices increase in the topology, the density calculation function reruns to get the current topology information (lines 2-4). After that, the management layer starts to function by performing the Q-learning in line-5. The Q-learning is performed by taking an action with the $\epsilon$-greedy approach and Q-table is updated according to the reward function results (lines 7-9). In the last step, the optimum action set is applied as the output of the autonomous recommendation engine (line-12).

\begin{algorithm}[ht]
  \caption{Q-CSM Algorithm}
  \label{alg::conjugateGradient}
  \begin{algorithmic}[1]
    \Require Number of smart city services (1-to-3), number of active IoT devices (m), KPI specifications, sensor asset set, operational asset set, service requests
    \Ensure QoS classes, QoS class density values, optimized actions to increase the average lifetime of IoT devices
    \State Initialize  $Q_i$, $O_i$;
    \State \textbf{foreach} change in $m$
    \State \ \ \ \ Calculate $\alpha(Q_i)$
    \State \textbf{end}
    \State \textbf{foreach} episode
    \State \ \ \ \ Take action, $a$
    \State \ \ \ \ Observe reward $R$, and state, $S'$
    \State \ \ \ \ Update Q-table 
    \State \ \ \ \ \ \ \
 \(Q_{t+1} \gets Q(.) + \alpha (R + \gamma \max_{A'} Q'-Q(.))\)
      \State \ \ \ \ Decide next action, $a'$ with policy, $\pi$ \
      \State \ \ \ \ Update $a$ $\gets$ $a'$, $S$ $\gets$ $S'$  \
      \State \ \ \ \ Update action set, $A_{t+1}$ $\gets$  $A_{t}$ 
     \State \textbf{end}
\end{algorithmic}
\end{algorithm}

\section{Performance Results}

In this part, we investigate the performance of our proposed framework in terms of (i) the efficiency in the Adaptation Layer by observing the response time of the IoT Agent manager according to the increasing number of active IoT sensors within the smart city network,  (ii) the efficiency in the Management Layer by observing the lifetime of IoT devices with the changing QoS levels and (iii) the cumulative reward value of the Q-learning algorithm in exploration and exploitation phases against changing learning rates. In our simulation, we work on 2 different smart city networks to increase the heterogeneity of the topology: 2-service and 3-service smart city scenarios. Here, the 2-service scenario refers to any two of the smart city services given in Table \ref{tab:table2} are implemented together within the same smart city network. On the contrary, in the 3-service scenario, all three services are applied within the single smart city network. We utilize these networks with one of the traditional methods \cite{protocoladaptive} in the current literature, as well as the proposed Q-CSM method. 

\begin{table}[thpb!]
    \centering
\caption{Simulation Parameters} 
\centering 
\begin{tabular}{l c} 
\hline 
Parameters&Values \\ [0.5ex]
\hline\hline 
Number of smart city services &  \{2, 3\} \\ 
Number of IoT sensors & \{10, 50, 98, 150\}\\
Number of episodes & 10000 \\
Learning rate  & \{0.7, 0.07, 0.007\}  \\
Discount factor  & 0.99 \\
Batch size & \{{32, 128, 256}\}\\
Update policy & Epsilon-greedy\\
Confidence interval  & 95\% \\
\hline 
\end{tabular}
\label{tab:tab3}
\end{table}

\textit{Experimental Setup:} 
We create a dynamic smart city scenario with three smart services by using Python and the MATLAB R2023a\textsuperscript{\copyright}. 
To communicate with the dynamic scenario, we implement IoT application layer protocols MQTT, CoAP and HTTP in Python scripts. The simulation parameters are given in Table \ref{tab:tab3}.

\begin{figure}[h]
\centerline{\includegraphics[width=.36\textwidth]{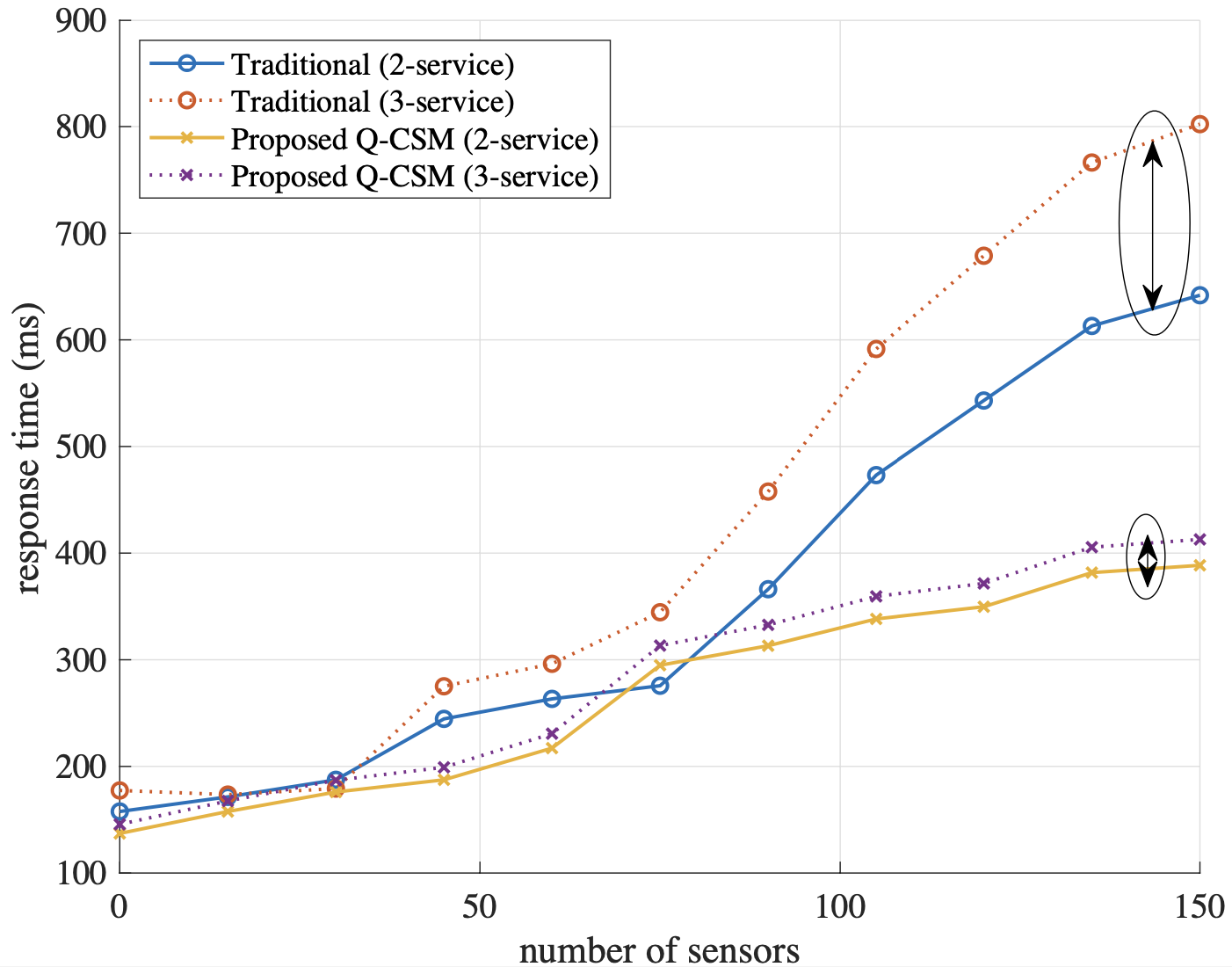}}
\caption{Response time comparison of IoT Agent Manager with the increasing number of active IoT devices.}
\label{fig1}
\end{figure}

We first explore the efficiency of the Adaptation Layer in Q-CSM. For this, we run 2-service and 3-service smart city scenarios, starting with the ten active IoT sensors within both. In addition, we assume that the city services utilize different IoT data protocols. For instance, if the wind turbine service works based on CoAP, then the other service(s) should not be based on CoAP; it should implement MQTT or HTTP. With this constraint, we maintain IoT data format heterogeneity to test the IoT Agent Manager. After that, by increasing this number up to hundred and fifty, we measure the response time of the IoT Agent Manager in ms while performing identical queries to the agent. Our recorded results are given in Fig. \ref{fig1}. According to the results, we note that our proposed Q-CSM framework responds to the queries by 38.7\% faster in the 2-service scenario. Even this fast response reaches $\sim$ 50\% levels in the 3-service scenario. As seen from the indicated circles in Fig. \ref{fig1}, the proposed Q-CSM method is not significantly affected by the heterogeneity level; the response times of the IoT Agent Manager are close to each other even when the number of sensors reaches high values. On the other hand, as in the traditional method, there is no data type conversion for the different IoT data protocols; this increases response time when the heterogeneity of the topology increases. 

\begin{figure}[h]
\centerline{\includegraphics[width=.36\textwidth]{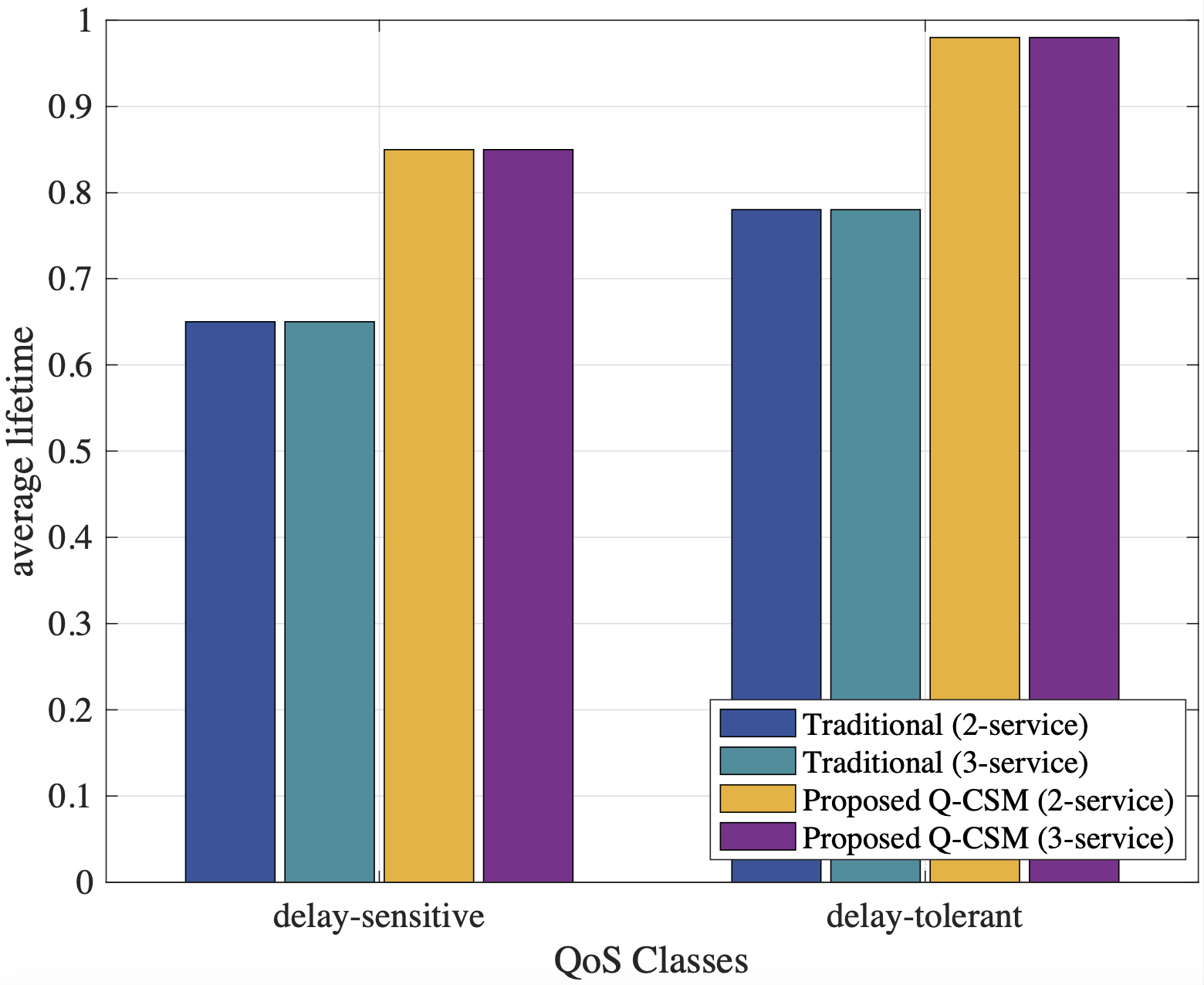}}
\caption{The average lifetime comparison of the IoT sensors against different QoS classes.}
\label{qos}
\end{figure}

In the performance investigation of the Management Layer, we observe the average lifetime of IoT devices depending on the two QoS classes: delay-sensitive and delay-tolerant. For this, we use the same 2-service and 3-service smart city networks with fifty active IoT sensors within the topology. Here, we assume the maximum value for the lifetime of IoT devices is 10 years, which decreases in proportion to the requested data. We run both methods separately for a twenty-minute simulation time and record the remaining battery values. After that, we form average lifetime values by performing normalization. As seen from Fig. \ref{qos}, the average lifetime value is unaffected by the number of services for both the traditional method and Q-CSM. However, with the same service scenario and the QoS class, the proposed Q-CSM results in a 19.8\% longer lifetime for the sensors. The main reason for this is the capability of Q-CSM to make decisions by considering the optimized lifetime depending on the desired IoT QoS classes. In addition, as we utilize Q-learning within the Q-CSM, we enhance the quality of decisions via the exploration and exploitation phases in the algorithm.

\begin{figure}[h]
\centerline{\includegraphics[width=.36\textwidth]{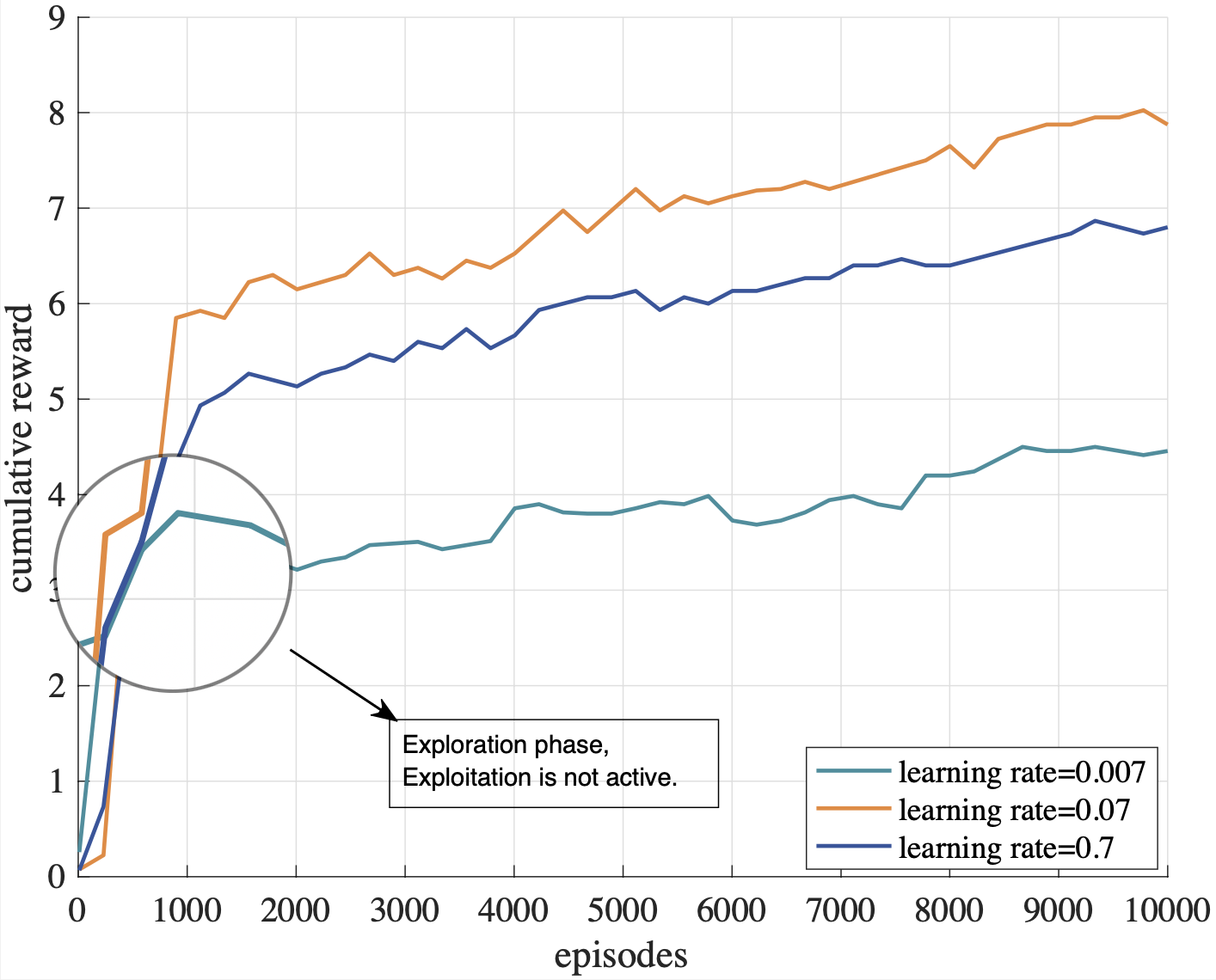}}
\caption{Cumulative reward value of Q-learning with changing learning rates during the simulation.}
\label{ql}
\end{figure}

Furthermore, we investigate the performance of the Q-learning algorithm by tracking the cumulative reward value. In this circumstance, we observe the training process consisting of \textit{exploration} and \textit{exploitation} phases. In addition, we run the algorithm for three different learning rates. In the exploration phase, the agent takes any action to discover new features regarding the environment. Conversely, in exploitation, the agent takes actions based on the knowledge gained. Based on these, we observe the training process against the episodes performed. In implementing Q-learning, we first deactivate the exploitation phase to increase the knowledge gained within the first 10\% of the episodes. That's why the cumulative reward value cannot reach its maximum value. We indicate this case with a zoomed circle in Fig. \ref{ql}. Furthermore, we change the learning rates and rerun the simıulation. During the three iterations with changing learning rates, we note the following results: the learning rate equal to $0.07$ converges to the highest reward value. Conversely, the learning rates $0.7$ and $0.007$ also converge to a reward value but cannot result in the maximized value. The main reason is that lower learning rates can sometimes improve generalization to unseen scenarios.

\section{Conclusion}

Intelligent and resource-efficient solutions are required to combat the heterogeneity challenge in IoT networks. With this motivation, we introduce a Q-learning-based Cognitive Service Management (Q-CSM) framework in this study. We first design an IoT Agent Manager to perform as a proxy for different data formats in IoT data protocols. Then, we implement Q-learning algorithm to produce insightful actions that meet the QoS requirements of each IoT network service. For this, we consider real-time service requests and per-cycle communication metrics, such as maximum delay and worst-case loss rate, as inputs to the cognitive module. We test our proposed framework on a smart city application consisting of three distinct services. Our simulation results show that Q-CSM achieves better in terms of minimizing the response time and maximizing the lifetime of the IoT sensors.

\section*{Acknowledgment}
This work was supported by The Scientific and Technological Research Council of Turkey (TUBITAK) 1515 Frontier R\&D Laboratories Support Program for BTS Advanced AI Hub: BTS Autonomous Networks and Data Innovation Lab. Project 5239903.

\bibliographystyle{IEEEtran}
\bibliography{IEEEabrv, whole}

\begin{thebibliography}{10}
\providecommand{\url}[1]{#1}
\csname url@samestyle\endcsname
\providecommand{\newblock}{\relax}
\providecommand{\bibinfo}[2]{#2}
\providecommand{\BIBentrySTDinterwordspacing}{\spaceskip=0pt\relax}
\providecommand{\BIBentryALTinterwordstretchfactor}{4}
\providecommand{\BIBentryALTinterwordspacing}{\spaceskip=\fontdimen2\font plus
\BIBentryALTinterwordstretchfactor\fontdimen3\font minus \fontdimen4\font\relax}
\providecommand{\BIBforeignlanguage}[2]{{%
\expandafter\ifx\csname l@#1\endcsname\relax
\typeout{** WARNING: IEEEtran.bst: No hyphenation pattern has been}%
\typeout{** loaded for the language `#1'. Using the pattern for}%
\typeout{** the default language instead.}%
\else
\language=\csname l@#1\endcsname
\fi
#2}}
\providecommand{\BIBdecl}{\relax}
\BIBdecl

\bibitem{survey2}
G.~K. Walia, M.~Kumar, and S.~S. Gill, ``Ai-empowered fog/edge resource management for iot applications: A comprehensive review, research challenges, and future perspectives,'' \emph{IEEE Communications Surveys \& Tutorials}, vol.~26, no.~1, pp. 619--669, 2024.

\bibitem{tgcn}
K.~Duran and B.~Canberk, ``Digital twin enriched green topology discovery for next generation core networks,'' \emph{IEEE Transactions on Green Communications and Networking}, vol.~7, no.~4, pp. 1946--1956, 2023.

\bibitem{intelligent}
S.~Skaperas, L.~Mamatas, and V.~Tsaoussidis, ``A link-quality anomaly detection framework for software-defined wireless mesh networks,'' \emph{IEEE Transactions on Machine Learning in Communications and Networking}, vol.~2, pp. 495--510, 2024.

\bibitem{rfc7228}
\BIBentryALTinterwordspacing
C.~Bormann, M.~Ersue, and A.~Keränen, ``{Terminology for Constrained-Node Networks},'' RFC 7228, May 2014. [Online]. Available: \url{https://www.rfc-editor.org/info/rfc7228}
\BIBentrySTDinterwordspacing

\bibitem{dtaas}
K.~Duran, E.~Ak, G.~Yurdakul, and B.~Canberk, ``6g-enabled dtaas (digital twin as a service) for decarbonized cities,'' in \emph{2023 IEEE International Conference on Communications Workshops (ICC Workshops)}, 2023, pp. 421--426.

\bibitem{bozkaya}
\BIBentryALTinterwordspacing
E.~Bozkaya, M.~Karatas, and L.~Eriskin, ``Chapter 1 - heterogeneous wireless sensor networks: Deployment strategies and coverage models,'' in \emph{Comprehensive Guide to Heterogeneous Networks}, K.~Ahuja, A.~Nayyar, and K.~Sharma, Eds.\hskip 1em plus 0.5em minus 0.4em\relax Academic Press, 2023, pp. 1--32. [Online]. Available: \url{https://www.sciencedirect.com/science/article/pii/B9780323905275000095}
\BIBentrySTDinterwordspacing

\bibitem{ccnc}
K.~Duran, B.~Karanlik, and B.~Canberk, ``Ai-driven partial topology discovery algorithm for broadband networks,'' in \emph{2021 IEEE 18th Annual Consumer Communications \& Networking Conference (CCNC)}, 2021, pp. 1--6.

\bibitem{contextaware}
M.~Ozturk, A.~I. Abubakar, R.~N.~B. Rais, M.~Jaber, S.~Hussain, and M.~A. Imran, ``Context-aware wireless connectivity and processing unit optimization for iot networks,'' \emph{IEEE Internet of Things Journal}, vol.~9, no.~17, pp. 16\,028--16\,043, 2022.

\bibitem{protocoladaptive}
L.~Mamatas, V.~Demiroglou, S.~Kalafatidis, S.~Skaperas, and V.~Tsaoussidis, ``Protocol-adaptive strategies for wireless mesh smart city networks,'' \emph{IEEE Network}, vol.~37, no.~2, pp. 136--143, 2023.

\bibitem{camad}
A.~G. Avran, E.~Ak, K.~Duran, G.~Yurdakul, and G.~Seçinti, ``Securing southbound interface in sdns: Utilizing support vector machines for openflow packet classification,'' in \emph{2023 IEEE 28th International Workshop on Computer Aided Modeling and Design of Communication Links and Networks (CAMAD)}, 2023, pp. 258--263.

\bibitem{access2}
I.~Chakour, C.~Daoui, M.~Baslam, B.~Sainz-De-Abajo, and B.~Garcia-Zapirain, ``Strategic bandwidth allocation for qos in iot gateway: Predicting future needs based on iot device habits,'' \emph{IEEE Access}, vol.~12, pp. 6590--6603, 2024.

\bibitem{icc24}
L.~V. Cakir, K.~Duran, C.~Thomson, M.~Broadbent, and B.~Canberk, ``Ai in energy digital twining: A reinforcement learning-based adaptive digital twin model for green cities,'' \emph{arXiv preprint arXiv:2401.16449}, 2024.

\bibitem{trconsumer}
K.~H.~K. Reddy, R.~S. Goswami, A.~K. Luhach, P.~Chatterjee, M.~Alnumay, and D.~S. Roy, ``Eflsm:- an intelligent resource manager for fog layer service management in smart cities,'' \emph{IEEE Transactions on Consumer Electronics}, vol.~70, no.~1, pp. 2281--2289, 2024.

\bibitem{wiley}
K.~Duran, B.~Karanlik, and B.~Canberk, ``Graph theoretical approach for automated ip lifecycle management in telco networks,'' \emph{Wiley Int J Network Mgmt}, vol.~31, no. 4, e2138, 2021.

\bibitem{schedular}
W.-B. Sun, J.~Xie, X.~Yang, L.~Wang, and W.-X. Meng, ``Efficient computation offloading and resource allocation scheme for opportunistic access fog-cloud computing networks,'' \emph{IEEE Transactions on Cognitive Communications and Networking}, vol.~9, no.~2, pp. 521--533, 2023.

\bibitem{access}
C.~Pandey, V.~Tiwari, A.~L. Imoize, C.-T. Li, C.-C. Lee, and D.~S. Roy, ``5gt-gan: Enhancing data augmentation for 5g-enabled mobile edge computing in smart cities,'' \emph{IEEE Access}, vol.~11, pp. 120\,983--120\,996, 2023.

\bibitem{IoT_journal}
D.~Wu, M.~Sun, P.~Zhang, Y.~Tu, Z.~Yang, and R.~Wang, ``Personalized secure demand-oriented data service toward edge-cloud collaborative iot,'' \emph{IEEE Internet of Things Journal}, vol.~10, no.~1, pp. 378--390, 2023.

\bibitem{survey}
K.~Fizza, A.~Banerjee, P.~P. Jayaraman, N.~Auluck, R.~Ranjan, K.~Mitra, and D.~Georgakopoulos, ``A survey on evaluating the quality of autonomic internet of things applications,'' \emph{IEEE Communications Surveys \& Tutorials}, vol.~25, no.~1, pp. 567--590, 2023.

\bibitem{aot}
K.~Duran, M.~Özdem, T.~Hoang, T.~Q. Duong, and B.~Canberk, ``Age of twin (aot): A new digital twin qualifier for 6g ecosystem,'' \emph{IEEE Internet of Things Magazine}, vol.~6, no.~4, pp. 138--143, 2023.

\bibitem{t6conf}
E.~Ak, K.~Duran, O.~A. Dobre, T.~Q. Duong, and B.~Canberk, ``T6conf: Digital twin networking framework for ipv6-enabled net-zero smart cities,'' \emph{IEEE Communications Magazine}, vol.~61, no.~3, pp. 36--42, 2023.

\bibitem{wifi}
E.~Ak and B.~Canberk, ``Fsc: Two-scale ai-driven fair sensitivity control for 802.11ax networks,'' in \emph{GLOBECOM 2020 - 2020 IEEE Global Communications Conference}, 2020, pp. 1--6.

\end{thebibliography}

\vspace{12pt}

\end{document}